\documentclass[prb,amsmath,amssymb,twocolumn,longbibliography]{revtex4-2}
\usepackage{graphicx}
\usepackage{dcolumn}
\usepackage{bm}
\usepackage{placeins}
\usepackage{rotating}
\usepackage{multirow}
\usepackage[dvipsnames,usenames]{xcolor}
\usepackage[english]{babel}
\usepackage{amssymb}
\usepackage{amsmath}
\usepackage[utf8]{inputenc}
\usepackage[normalem]{ulem}
\usepackage{float}
\usepackage{siunitx}

\newcommand{\be}{\begin{equation}}
\newcommand{\ee}{\end{equation}}
\newcommand{\bea}{\begin{eqnarray}}
\newcommand{\eea}{\end{eqnarray}}
\newcommand{\nn} {\nonumber}

\def\a{\alpha}

\def\g{\gamma}
\def\G{\Gamma}
\def\d{\delta}
\def\D{\Delta}

\def\ve{\varepsilon}

\def\vf{\varphi}

\def\ket{\rangle}

\def\x{{\rm x}}

\graphicspath{{Figures/}}

\begin{document}
\widetext
\title{Accurate and efficient calculation of atomic forces in solids with \\ non-self-consistent hybrid functionals}
\author{Damian Contant}
\altaffiliation{Present address: Department of Physics, King’s College London, Strand campus, London WC2R 2LS, United Kingdom}
\affiliation{Sorbonne Universit\'e, MNHN, UMR CNRS 7590, IMPMC, 4 place Jussieu, 75005 Paris, France}
\author{Maria Hellgren}
\affiliation{Sorbonne Universit\'e, MNHN, UMR CNRS 7590, IMPMC, 4 place Jussieu, 75005 Paris, France}
\date{\today}

\begin{abstract}
Hybrid functionals are routinely employed self-consistently within the generalized Kohn-Sham framework. The evaluation of the nonlocal Fock exchange operator makes hybrid functional calculations computationally expensive, in particular with plane-wave basis sets. Here, we investigate the advantages of non-self-consistent hybrid functional calculations, focusing on the evaluation of atomic forces.
The analytical force terms that arise due to non-self-consistency are computed using density functional perturbation theory (DFPT), as implemented within the \verb|Quantum ESPRESSO| distribution. A non-self-consistent hybrid force calculation thus consists of self-consistent DFPT calculations with a local or semi-local functional and a single evaluation of the Fock exchange operator.
The overall computational cost is, thereby, reduced in general, especially for solids that require a dense Brillouin-zone sampling. Moreover, results for structural parameters are barely affected by self-consistency, and vibrational frequencies are typically agreeing within 0.5\%, thus making non-self-consistent calculations an interesting alternative.
\end{abstract}
\maketitle

\section{Introduction}

Hybrid functionals have a long-standing place in computational chemistry and are becoming more and more popular in materials science. The inclusion of a fraction of exact exchange within a generalized gradient approximation (GGA) drastically improves structural and electronic properties ~\cite{becke1993,becke1993_2,perdew1996,adamo1999ReliableDensityFunctional,ernzerhof1999}. One of the main reasons for the improvement is the inclusion of a realistic derivative discontinuity through nonlocal exchange \cite{perdew1982,stein2012,hellgren2012,hellgren2013}, a feature challenging to capture with semi-local \cite{Aschebrock2017} and meta-GGA functionals \cite{Eich2014,Aschebrock2019}.

There are, however, two main drawbacks. The first is that the optimal fraction of exact exchange is system dependent, while for most hybrid functionals such as, for example, the PBE0 \cite{adamo1999ReliableDensityFunctional,ernzerhof1999} and HSE \cite{heyd2003HybridFunctionalsBased,heyd2006ErratumHybridFunctionals}, a default fraction is set for all types of systems. Various ways to refine the fraction of exact exchange have, therefore, been proposed \cite{skone2014,skone2016,hellgren2021,pitts2025SelfconsistentRandomPhase}.
The second drawback is the computational cost. Hybrid functionals intrinsically scale as $N_e^4$/$N_k^2$ with the number of electrons/$k$-points. A hybrid functional calculation is, therefore, typically two orders of magnitude more expensive than, e.g., a semi-local PBE calculation. In this regard, there has been a number of developments to reduce the scaling and improve efficiency, notably the adaptive compressed exchange (ACE) operator method, which reduces the total number of evaluations of the nonlocal Fock operator ~\cite{lin2016}.

So far, almost all hybrid calculations are carried out self-consistently. This means that the ground-state energy is optimized together with the electronic density. The latter can change significantly in strongly correlated systems or close to an electronic phase transition. On the other hand, many interesting problems deal only with the ground-state energy and small variations of this energy around a stable local minimum. For these situations, it is motivated to use a non-self-consistent approach, which is already the standard procedure for advanced functionals such as MP2 \cite{10.1063/1.4919238} and RPA \cite{burow2014AnalyticalFirstOrderMolecular,ramberger2017AnalyticInteratomicForces,contant2026}. The accuracy obtained in such calculations for hybrid functionals was studied in Ref.~\cite{skelton2020}, showing that band gaps were, in fact, not very sensitive to self-consistency. Using a PBE starting point, the differences observed were within 5\%, with exceptions for strongly correlated systems.

In this work, we investigate the effects of self-consistency on the atomic forces, with the aim of reducing the computational cost of hybrid functionals. Accurate forces are important for several types of calculations related to structural and vibrational properties \cite{baroni2001PhononsRelatedCrystal}. Such calculations are typically very expensive and, therefore, mostly rely on semi-local approximations, which do not always achieve sufficient accuracy \cite{franchini2005,Franchini2014,casadei2012,drummond2015quantum,Gillan2016,hellgrentise2_2017,gruber2018,hellgren2021RandomPhaseApproximation,fransson2023}.

Contrary to the evaluation of band gaps, non-self-consistent forces require the implementation of additional analytical terms. These extra contributions involve the calculation of self-consistent orbital responses with respect to atomic displacements, which adds computational time. Nevertheless, as will be demonstrated in this work, the cost of evaluating these terms is, in most cases, lower than the repeated evaluation of the Fock exchange operator, as in the fully self-consistent procedure.

The paper is organized as follows. In Sec.~II, we present the equations for evaluating non-self-consistent forces with hybrid functionals and discuss their implementation. In Sec. III, we perform several test studies on molecular crystals (H$_2$O and H$_2$), bulk (C, Si, Ge, SiO$_2$, and TiO$_2$) and 2D materials (graphene, silicene, boron nitride, and phosphorene). We discuss the accuracy and computational cost of the non-self-consistent calculations with respect to standard self-consistent calculations done with the ACE method. Finally, in Sec. IV, we present our conclusions.

\section{Forces with non-self-consistent hybrid functionals}

In this Section, we present the equations for evaluating atomic forces with non-self-consistent hybrid functionals. Although these expressions are valid for any functional that mixes in a fraction of exact exchange into a semi-local functional, we will here focus on the PBE0 hybrid functional. The PBE0 exchange-correlation (xc) energy is given by
 \begin{equation}
E^{\rm PBE0}_{\text{xc}}=\a E^{\rm HF}_\x + (1-\a)E^{\rm PBE}_{\rm x}+E^{\rm PBE}_{\rm c},
\label{pbe0}
\end{equation}
where $E^{\rm HF}_\x$ is the exact-exchange energy, as calculated within the Hartree-Fock approximation. The parameter $\a$ can be varied but is set to 25\% by default, following Refs.~\cite{perdew1996,adamo1999ReliableDensityFunctional,ernzerhof1999}.

The PBE0 approximation is an explicit functional of the electronic first-order reduced density matrix $\g(\mathbf{r},\mathbf{r}')$. Allowing free variations of $\g$ under the constraint that $\g$ comes from a single Slater determinant leads to a set of single particle equations with an effective nonlocal potential \cite{parryang}. The latter consists of the external potential, $v_{\rm ext}(\mathbf{r})$, the Hartree potential, $v_{\rm H}(\mathbf{r})$, and the xc potential
\begin{equation}
V_{\rm xc}^{\rm PBE0}(\mathbf{r},\mathbf{r}')=\a V_\x(\mathbf{r},\mathbf{r}') + (1-\a)v^{\rm PBE}_{\rm x}(\mathbf{r})+v^{\rm PBE}_{\rm c}(\mathbf{r}),
\label{pbe0_pot}
\end{equation}
where the nonlocal component, $V_\x$, is the exact-exchange potential of the HF approximation.

In a self-consistent (scf) calculation, the atomic forces have a simple expression, thanks to the well-known Hellmann-Feynman theorem \cite{hellmann1937EinfuehrungQuantenchemie,Feynman1939}
\begin{equation}
\begin{aligned}
\displaystyle \mathbf{F}_{I} = & - \frac{\partial E_{\rm NN}}{\partial \mathbf{R}_{I}} - \int \! \g_{\rm scf}(\mathbf{r},\mathbf{r}') \frac{\partial v_{\text{ext}}(\mathbf{r},\mathbf{r}')}{\partial \mathbf{R}_{I}} \, d\mathbf{r} d\mathbf{r}'.
\end{aligned}
\label{hfforce}
\end{equation}
The nuclear-nuclear (NN) potential energy is given by
\be
E_{\rm NN}=\sum \limits_{A \neq I} \frac{Z_{A} Z_{I}}{|\mathbf{R}_{A} - \mathbf{R}_{I}|},
\label{nnenergy}
\ee
with $Z_{I}$ the nuclear charge of atom $I$ at position $\mathbf{R}_{I}$, and the external electron-nuclear potential is given by
\begin{equation}
v_{\text{ext}}(\mathbf{r},\mathbf{r}') = - \sum \limits_{A} \frac{Z_{A}}{|\mathbf{R}_{A} - \mathbf{r}|}\d (\mathbf{r},\mathbf{r}').
\label{extenergy}
\end{equation}
In the pseudopotential approximation \cite{kleinman1982EfficaciousFormModel}, the external frozen-core potential is, in general, constructed such that it contains both a local (L) and a nonlocal (NL) component,
\begin{equation}
v_{\text{ext}}(\mathbf{r},\mathbf{r}')\approx v_{\rm L}(\mathbf{r})\d (\mathbf{r},\mathbf{r}')+v_{\rm NL}(\mathbf{r},\mathbf{r}').
\label{extpot}
\end{equation}
The nonlocal pseudopotential component needs to be treated with care, in particular when applying the Hellmann-Feynman expression (Eq.~(\ref{hfforce})) within the self-consistent optimized effective potential (OEP) method \cite{contant2024OptimizedEffectivePotential}.

It is also possible to evaluate forces starting from a set of orbitals different from the self-consistent ones. Within PBE0, a natural choice would be the self-consistent PBE orbitals. In this case, the expression for the forces, given in Eq.~(\ref{hfforce}), has to be modified according to
\begin{eqnarray}
\mathbf{F}_{I}& = & - \frac{\partial E_{\rm NN}}{\partial \mathbf{R}_{I}} - \int \! \g^{\rm PBE}(\mathbf{r},\mathbf{r}') \frac{\partial v_{\text{ext}}(\mathbf{r},\mathbf{r}')}{\partial \mathbf{R}_{I}} \, d\mathbf{r} d\mathbf{r} '\nn \\
& &+ \int\! v^{\rm PBE}_{\text{xc}}(\mathbf{r}) \frac{\partial n^{\rm PBE}(\mathbf{r})}{\partial \mathbf{R}_{I}} \, d\mathbf{r} - \frac{\partial E^{\rm PBE0}_{\text{xc}}}{\partial \mathbf{R}_{I}}.
\label{nscfforce}
\end{eqnarray}
The two additional terms that arise in such non-self-consistent (nscf) calculations can be grouped into the following force correction term
\begin{eqnarray}
\!\!\!\Delta \mathbf{F}_{\text{xc}}^{\text{nscf}} \nn\\
&& \!\!\!\!\!\!\!\!\!\!\!\!\!\!\!\!\!\!\!\!=\int \!v^{\text{PBE}}_{\text{xc}}(\mathbf{r}) \frac{\delta n^{\text{PBE}}(\mathbf{r})}{\delta v_{\text{eff}}(\mathbf{r}',\mathbf{r}'')} \frac{\partial v_{\text{eff}}(\mathbf{r}',\mathbf{r}'')}{\partial \mathbf{R}_{I}} \, d\mathbf{r} \, d\mathbf{r}' \, d\mathbf{r}''\nn\\
&& \!\!\!\!\!\!\!\!\!\!\!\!\!\!\!\! -\int \! \frac{\delta E^{\rm PBE0}_{\text{xc}}}{\delta v_{\text{eff}}(\mathbf{r},\mathbf{r}')} \frac{\partial v_{\text{eff}}(\mathbf{r},\mathbf{r}')}{\partial \mathbf{R}_{I}} \, d\mathbf{r} d\mathbf{r}' ,
\label{nscfforceterm}
\end{eqnarray}
where the derivatives with respect to nuclear coordinates have been expanded using the chain rule over the effective PBE KS potential. The derivative of the latter with respect to nuclear coordinates is given by
\begin{eqnarray}
\frac{\partial v_{\text{eff}}(\mathbf{r},\mathbf{r}')}{\partial \mathbf{R}_{I}} &=&\left\{\frac{\partial v_{\text{H}}(\mathbf{r})}{\partial \mathbf{R}_{I}}+\frac{\partial v^{\rm PBE}_{\text{xc}}(\mathbf{r})}{\partial \mathbf{R}_{I}}+\frac{\partial v_{\text{L}}(\mathbf{r})}{\partial \mathbf{R}_{I}}\right\}\nn\\
&&\times\delta(\mathbf{r}-\mathbf{r'})+\frac{\partial v_{\text{NL}}(\mathbf{r},\mathbf{r}')}{\partial \mathbf{R}_{I}}.
\label{effpotder}
\end{eqnarray}
The functional derivative of the PBE0 xc energy with respect to $v_{\rm eff}$ can be further expanded in terms of the density matrix to yield
\begin{eqnarray}
\!\!\!\Delta \mathbf{F}_{\text{xc}}^{\text{nscf}} &&\nn\\
&&\!\!\!\!\!\!\!\!\!\!\!\!\!\!\!\!\!\!\!\!=\a\int \! v^{\text{PBE}}_{\text{x}}(\mathbf{r}) \frac{\delta n^{\text{PBE}}(\mathbf{r})}{\delta v_{\text{eff}}(\mathbf{r}',\mathbf{r}'')} \frac{\partial v_{\text{eff}}(\mathbf{r}',\mathbf{r}'')}{\partial \mathbf{R}_{I}} \, d\mathbf{r} \, d\mathbf{r}' \, d\mathbf{r}''\nn\\
&&\!\!\!\!\!\!\!\!\!\!\!\!\!\!\!\! -\a\int \! V_{\text{x}}(\mathbf{r},\mathbf{r}')\frac{\delta \g^{\rm PBE}(\mathbf{r},\mathbf{r}')}{\d v_{\text{eff}}(\mathbf{r}'',\mathbf{r}''')} \frac{\partial v_{\text{eff}}(\mathbf{r''},\mathbf{r}''')}{\partial \mathbf{R}_{I}} \, \nn\\
&&\!\!\!\!\!\times d\mathbf{r} d\mathbf{r}'d\mathbf{r}''d\mathbf{r}''' ,
\label{nscfforcetermexpand}
\end{eqnarray}
where the local xc contributions have been canceled due to the fact that PBE0 and PBE share some of the xc terms. Such cancellations will not occur with other starting points.

\begin{figure}[t]
\centering
\includegraphics[scale=0.49]{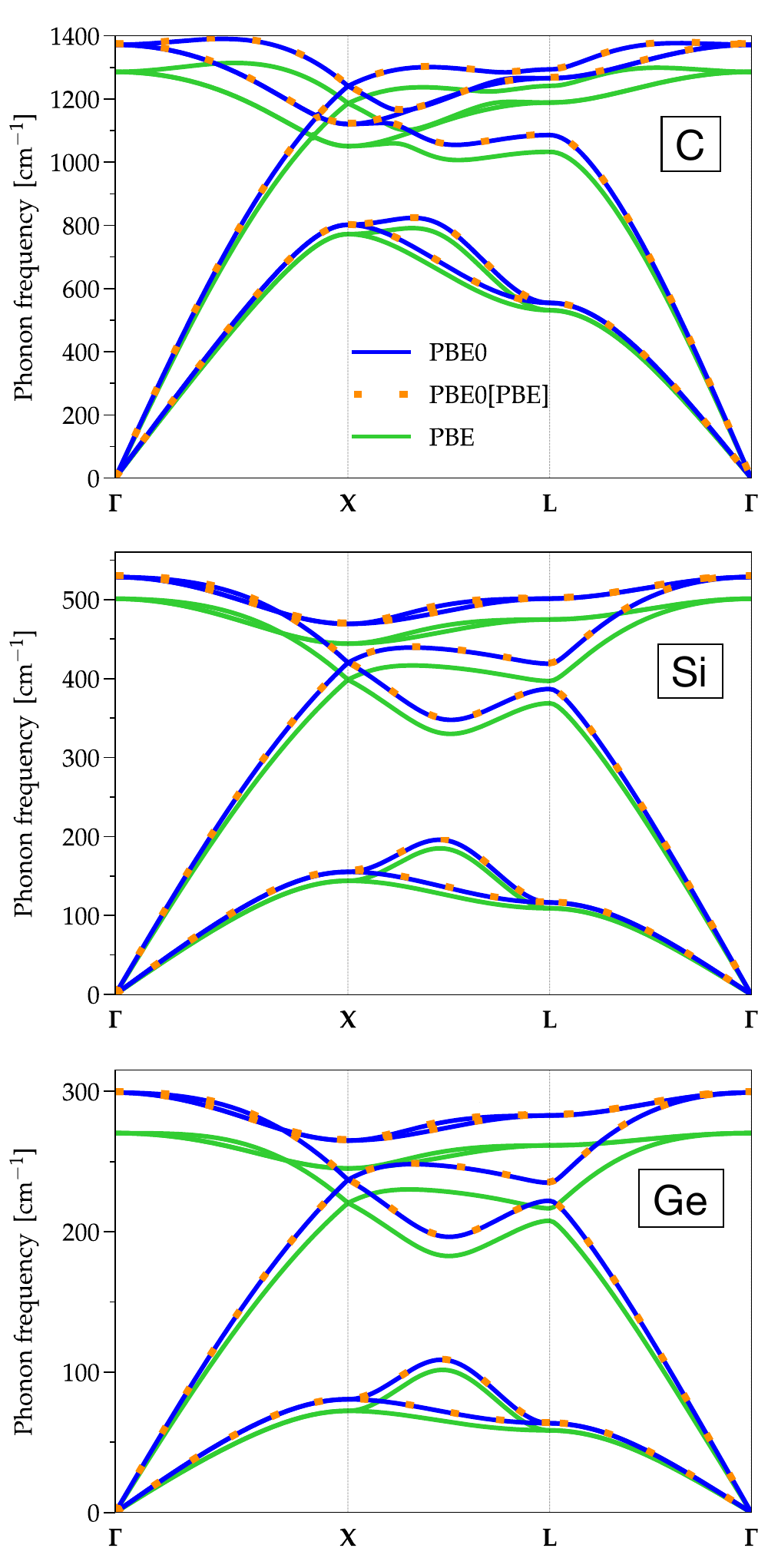}
\caption{Phonon dispersion of C, Si, and Ge, calculated with PBE0 (blue), PBE0[PBE] (orange dots), and PBE (green), on top of their respective optimal structure (see Tab.~\ref{tab:C_Si_Ge_geometry})
}\label{fig:diamond_nscf}
\end{figure}

To evaluate Eq.~(\ref{nscfforcetermexpand}), we need the self-consistent PBE orbital responses with respect to perturbations induced by atomic displacements. Such responses can be calculated through the framework of density functional perturbation theory (DFPT), in which the orbital responses are evaluated within the self-consistent Sternheimer equation \cite{sternheimer1954ElectronicPolarizabilitiesIons,baroni2001PhononsRelatedCrystal}
\be
\left [ H_{\rm KS}+\g P_{\rm occ}-\ve_{n}\right ]|\D\vf_{n}\ket=-(1-P_{\rm occ})\D v_{\rm eff}|\vf_{n}\ket.
\label{imp:lin2}
\ee
Here, $\hat{H}_{\rm KS}$ is the unperturbed PBE KS Hamiltonian, $\varepsilon_{n}$/$\varphi_{n}$ the corresponding occupied KS eigenvalue/orbital, and $\Delta \varphi_{n}$ the first-order orbital response to the applied perturbation contained in $\D v_{\rm eff}$. The operator $P_{\rm occ}$ projects onto the occupied state manifold, which is sufficient to evaluate Eq.~(\ref{nscfforcetermexpand}). This quantity also ensures that the linear system is non-singular on the left-hand side of the Sternheimer equation \cite{baroni2001PhononsRelatedCrystal,sternheimer1954ElectronicPolarizabilitiesIons}.

Non-self-consistent forces at the PBE0 level have, in this work, been implemented within the plane-wave and pseudopotential framework, and incorporated into the \verb|ACFDT| package \cite{nguyen2009,nguyen2014InitioSelfconsistentTotalenergy,colonna2014CorrelationEnergyExactexchange,Colonna2016,hellgren2018RandomPhaseApproximation,hellgren2021RandomPhaseApproximation,pitts2025SelfconsistentRandomPhase,contant2024OptimizedEffectivePotential,contant2026} of the \verb|Quantum ESPRESSO| (\verb|QE|) distribution \cite{Giannozzi2009,Giannozzi2017,Giannozzi2020}. In the \verb|ACFDT| package, both molecules and solids can be treated using norm-conserving pseudopotentials. The extra force term given in Eq.~(\ref{nscfforcetermexpand}) is evaluated using DFPT routines available in the \verb|PHonon| package of \verb|QE|. For every $k$-point, a single evaluation of the nonlocal exchange operator is required. The latter is used for determining the second term in Eq.~(\ref{nscfforcetermexpand}) and for computing the total nscf energy.

With a local or a semi-local approximation, the DFPT step scales linearly with the number of $k$-points, while the nonlocal exchange operator instead has a quadratic scaling. With a moderately dense $k$-point sampling, the cost of multiple evaluations of the exchange operator easily exceeds the cost of solving the self-consistent DFPT equations, as long as the total number of perturbations to calculate is not too large. The latter is proportional to the number of atoms in the system.

In the next Section, we will look at various examples showing that, even compared to the efficient ACE implementation, we gain in speed and computational cost, without loss of relevant accuracy.

\section{Applications}

In this Section, we present the calculations done on molecular crystals (H$_2$ and H$_2$O), a set of bulk materials (C, Si, Ge, SiO$_2$, and TiO$_2$), and 2D materials (graphene, silicene, boron nitride, and phosphorene). We compare nscf PBE0 hybrid functional calculations starting from PBE, from now on denoted PBE0[PBE], to scf PBE0 calculations within the ACE implementation \cite{lin2016} (default option in the \verb|PW| package of the \verb|QE| distribution). PBE optimized norm-conserving Vanderbilt (ONCV) pseudopotentials \cite{hamann2013OptimizedNormconservingVanderbilt} are employed throughout, unless otherwise stated. All total energy and force calculations are done with the \verb|PW| and \verb|ACFDT| packages of the \verb|QE| distribution. Phonon frequencies are subsequently obtained via finite differences of forces using the \verb|Phonopy| code \cite{togo2023FirstprinciplesPhononCalculations,togo2023ImplementationStrategiesPhonopy}. We note that cell optimization based on the calculation of stresses is currently not fully implemented within \verb|QE| for hybrid functionals. As a consequence, the optimal lattice parameters are obtained by manually minimizing the total energy.

\begin{table*}[t]
\caption{\label{tab:C_Si_Ge_geometry} Lattice parameter ($a$) [\AA] and $\Gamma$-point optical phonon frequency (${\nu}$) [cm$^{-1}$] calculated for C, Si, and Ge in PBE, PBE0, and PBE0[PBE]. For comparison, experimental lattice parameters are provided (C: \cite{occelli2003PropertiesDiamondHydrostatic}, Si: \cite{kittel2005IntroductionSolidState}, Ge: \cite{straumanis1952LatticeParametersCoefficients}), as well as experimental results corrected (corr) to account for zero-point anharmonic effects \cite{harl2010AssessingQualityRandom}. Experimental phonon frequencies are also listed (C: \cite{occelli2003PropertiesDiamondHydrostatic}, Si: \cite{Kulda1994}, Ge: \cite{nilsson1971PhononDispersionRelations}), as well as the corrected phonon frequencies obtained by subtracting anharmonic effects, as calculated in Ref.~\cite{vanderbilt1986CalculationAnharmonicPhonon}.
}
\centering
\setlength{\tabcolsep}{8pt}
\begin{tabular}{l|cc|cc|cc}
\hline
\hline
    \textbf{Solid}  &  \multicolumn{2}{c|}{\textbf{C}} & \multicolumn{2}{c|}{\textbf{Si}} & \multicolumn{2}{c}{\textbf{Ge}} \\
\hline
    \textbf{[\AA]/[cm$^{-1}$] }  &   $a$  & $\nu$ &   $a$  & $\nu$ &   $a$  & $\nu$ \\
\hline
    \textbf{PBE}                 &  3.569 &  1285 &  5.478 &  501  &  5.775 &  270  \\
    \textbf{PBE0}                &  3.544 &  1371 &  5.445 &  529  &  5.707 &  299  \\
    \textbf{PBE0[PBE]}           &  3.543 &  1374 &  5.444 &  530  &  5.705 &  300  \\
\hline
    \textit{References}          &        &       &        &       &        &       \\
    \textbf{Expt.}               &  3.567 &  1334 &  5.430 &  523  &  5.658 &  304  \\
    \textbf{Expt. (corr)}        &  3.553 &  1351 &  5.421 &  527  &  5.644 &  305  \\
\hline
\end{tabular}
\end{table*}

\subsection{C, Si, and Ge}

We first consider the diamond phase of carbon, silicon, and germanium ($Fd\overline{3}m$, space group no.~\!227), whose primitive unit cell contains two atoms, with positions fixed by symmetry. The lattice constant, $a$, and the phonon dispersion are computed with PBE, PBE0, and PBE0[PBE].

A plane-wave cutoff of 100~Ry ensures converged results for all materials. For the Brillouin-zone sampling, we used, for C and Si, a regular $\G$-centered 8$\times$8$\times$8 $k$-point grid for all three levels of approximations. For Ge, however, we used a 6$\times$6$\times$6 shifted $k$-point grid for the two hybrid functionals, and a 12$\times$12$\times$12 shifted grid for PBE. The Ge 14 electron ONCV pseudopotential was also replaced by a 4 electron Hartwigsen-Goedecker-Hutter (HGH) pseudopotential \cite{goedecker1996SeparableDualspaceGaussian,hartwigsen1998RelativisticSeparableDualspace}.
The results can be found in Tab.~\ref{tab:C_Si_Ge_geometry}.

As expected, PBE0 leads to more compact structures as compared to PBE, in better agreement with experimental results \cite{occelli2003PropertiesDiamondHydrostatic,kittel2005IntroductionSolidState,straumanis1952LatticeParametersCoefficients} (corrected by zero-point anharmonic effects \cite{harl2010AssessingQualityRandom}). The effect of self-consistency within PBE0 is limited to a maximum of 0.002~\!\AA. The fact that the PBE ground state is semi-metallic in Ge, while gapped in PBE0, does not seem to have a relevant impact on the computed PBE0[PBE] lattice constant.

Thanks to symmetry, the forces only need to be calculated on a single distorted geometry, in order to construct the entire dynamical matrix. The diamond phase has 6 normal modes: three acoustic and three optical. At the Brillouin-zone center ($\Gamma$-point), the three optical modes are all degenerate. The values for each material, computed in every approximation on their respective optimal structure, are presented in Tab.~\ref{tab:C_Si_Ge_geometry}. The agreement between PBE0 and PBE0\text{[}PBE\text{]} is excellent on all three systems. Compared to experimental results \cite{occelli2003PropertiesDiamondHydrostatic,Kulda1994,nilsson1971PhononDispersionRelations} corrected for anharmonic effects at the experimental lattice parameter \cite{vanderbilt1986CalculationAnharmonicPhonon}, PBE0 makes an important correction to PBE (e.g., about 100 cm$^{-1}$ (or 7\%) in C). The error of PBE0 with respect to experiment can be shown to be mostly due to the differences between the experimental and the optimized lattice constant \cite{contant2026}.
 
The phonon dispersion has been calculated on a [2,2,2] supercell (16 atoms). This gives access to an exact evaluation of the vibrational frequencies at the $X$-point (0.5~0.5~0.0) and $L$-point (0.5~0.5~0.5), in addition to $\G$. The points located in-between are obtained by Fourier interpolation. In accordance with the size of the supercell, the ${k}$-point grid used for C and Si was reduced to 4$\times$4$\times$4 for these calculations. For Ge, we used a 3$\times$3$\times$3 and a 6$\times$6$\times$6 sampling for the two hybrid functionals and PBE, respectively. The results are presented in Fig.~\ref{fig:diamond_nscf}. The quality of the nscf frequencies is found to be good not only at the $\Gamma$-point, but also along the full phonon dispersion. 
For C, the maximum frequency difference seen at one of the three high-symmetry points is only 3~cm$^{-1}$ between nscf and scf PBE0. For Si, the maximum difference is 2~cm$^{-1}$, and, for Ge, within 1~cm$^{-1}$. In the case of PBE0, the use of PBE starting orbitals is thus perfectly appropriate on all three systems, yielding lattice constants and phonon frequencies that essentially coincide with the self-consistent PBE0 solution.

When running these calculations on 32 cores on a local machine, an important speedup is noticed. The PBE0[PBE] evaluation of forces is found to be around 7 times faster on the primitive unit cell compared to scf PBE0, and about 2 times faster on the [2,2,2] supercell. 

We note that the speedup observed can vary in magnitude depending on how the different convergence thresholds are set, as well as the parallelization strategy adopted, and the hardware used for the calculations. Nevertheless, having tried different machines and setups, the numbers given here, and henceforth, can be seen as representative for a typical calculation.

\subsection{Ice}

\begin{table*}[t]
\caption{\label{tab:ice_geometry_phonons} Intramolecular O-H bond distance ($d_{\rm O-H}$) [\AA], intramolecular H-O-H bond angle ($\Theta_{\rm H-O-H}$) [$^{\circ}$], hydrogen bond distance ($D_{\rm H-O}$) [\AA], and highest vibrational frequency at the $\Gamma$-point ($\nu_{\rm O-H}$) [cm$^{-1}$] for phases VIII and XI of solid water (ice), calculated with PBE, PBE0, and PBE0[PBE]. The MA\%E of all optical vibrational frequencies ($\nu$) is also given with respect to scf PBE0. The experimental unit cell parameters have been used in all calculations \cite{kuhs1984StructureHydrogenOrdering,leadbetter1985TheEquilibriumLowTemp}.
}
\setlength{\tabcolsep}{6.8pt}
\begin{tabular}{l|ccccc|ccccc}
\hline
\hline
    \textbf{Solid}  &  \multicolumn{5}{c|}{\textbf{VIII}} & \multicolumn{5}{c}{\textbf{XI}} \\
\hline
    \textbf{[\AA]/[$^{\circ}$]}  &  $d_{\rm O-H}$ & $\Theta_{\rm H-O-H}$ & $D_{\rm H-O}$ & $\nu_{\rm O-H}$ & MA\%E ($\nu$) & $d_{\rm O-H}$ & $\Theta_{\rm H-O-H}$ & $D_{\rm H-O}$ & $\nu_{\rm O-H}$ & MA\%E ($\nu$) \\
\hline
    \textbf{PBE}        &  0.9872 & 105.43 & 1.8890 & 3441 & 3.89 & 0.9964 & 106.22 & 1.7584 & 3322 & 4.56 \\
    \textbf{PBE0}       &  0.9731 & 105.86 & 1.9029 & 3632 & 0.00 & 0.9803 & 106.50 & 1.7754 & 3538 & 0.00 \\
    \textbf{PBE0[PBE]}  &  0.9725 & 105.84 & 1.9035 & 3641 & 0.30 & 0.9798 & 106.54 & 1.7750 & 3547 & 0.22 \\
\hline
\end{tabular}
\end{table*}

Next, we look at molecular crystals. We first consider two different phases of solid water (ice) \cite{chaplin2019StructurePropertiesWater}. Phase~XI is a high-symmetry variant ($Cmc2_1$, space group no.~\!36 \cite{leadbetter1985TheEquilibriumLowTemp}) of the ordinary proton-disordered ice. The lattice is hexagonal and the primitive cell contains 4 water molecules, arranged in hydrogen-bonded configurations. Phase~VIII has almost half the volume of phase~XI and is formed at low temperature and under intermediate pressures. This phase belongs to the tetragonal crystal system ($I4_{1}/amd$, space group no.~\!141 \cite{kuhs1984StructureHydrogenOrdering}). The primitive cell of ice~VIII also contains 4 water molecules, but arranged in both hydrogen-bonded and non-hydrogen-bonded configurations.

The atomic positions that form the unit cell geometry have all been relaxed using the Broyden-Fletcher-Goldfarb-Shanno (BFGS) \cite{broyden1970ConvergenceClassDoublerank,fletcher1970NewApproachVariable,goldfarb1970FamilyVariablemetricMethods,shanno1970ConditioningQuasiNewtonMethods} algorithm, applied to PBE, PBE0, and PBE0[PBE]. This relaxation, which minimizes the total energy and the atomic forces, has been done at fixed lattice parameters, set to their experimental values (phase~XI (primitive cell parameters): $a\!=\! 4.502$~\!\AA, \, $c\!=\!7.328$~\!\AA~\cite{leadbetter1985TheEquilibriumLowTemp}, and phase~VIII (conventional cell parameters): $a\!=\! 4.65$~\!\AA, \, $c\!=\!6.77$~\!\AA~\cite{kuhs1984StructureHydrogenOrdering}). The $\G$-point phonon frequencies have then been calculated for each method on their corresponding relaxed geometry.
Converged results are obtained using a plane-wave kinetic energy cutoff of 150~Ry, and a 5$\times$5$\times$5/6$\times$6$\times$6 uniform $k$-point grid for ice XI/VIII, respectively. The same set-up is used for every method considered. Results are presented in Tab.~\ref{tab:ice_geometry_phonons}.

As expected, we find a smaller O-H intramolecular bond distance, $d_{\rm O-H}$, with PBE0 as compared to PBE. Furthermore, the intramolecular bond angle, $\Theta_{\rm H-O-H}$, widens and the hydrogen bond distance, $D_{\rm H-O}$, increases with PBE0 as compared to PBE. For these two systems, self-consistency does not appear relevant. The geometries obtained using either PBE0 or PBE0[PBE] are close to identical. The difference in bond distances for ice~VIII are within 0.0006~\!\AA, and the bond angles within 0.02$^{\circ}$. The same trend applies to ice~XI.

The $\G$-point vibrational frequencies have been calculated for both phases, leading to a total of 36 normal modes for each. However, unlike the systems considered so far, ice is a polar material. As such, the longitudinal optical (LO) frequencies are expected to be shifted with respect to the transverse optical (TO) modes due to long-range electrostatic effects \cite{zhong1994GiantLOTOSplittings,gonze1997DynamicalMatricesBorna}. For our study, however, since we are only interested in comparisons between self-consistent and non-self-consistent PBE0, we have not calculated the frequency shift associated with the LO-TO splitting.

In Tab.~\ref{tab:ice_geometry_phonons}, we present the mean absolute percentage error (MA\%E) of all optical vibrational modes, $\nu$, in PBE and PBE0[PBE], with respect to PBE0. The agreement between PBE0[PBE] and PBE0 is excellent, with an error of 0.30/0.22\% in phase VIII/XI, respectively. For the modes above 3000~cm$^{-1}$, which correspond to the stretching of O-H intramolecular bonds, the maximum difference observed is 11~cm$^{-1}$. The highest of these modes, $\nu_{\rm O-H}$, is presented in Tab.~\ref{tab:ice_geometry_phonons}. We can compare the difference between PBE0[PBE] and scf PBE0 to the difference between PBE and scf PBE0. In the latter case, the MA\%E is as large as 4\% and the difference in O-H intramolecular vibrational frequency close to 200~cm$^{-1}$.

In terms of computational cost, when ran on 32 cores on a supercomputer, the non-self-consistent calculations performed on the distorted unit cells of ice XI were found to be about 4 times faster than the self-consistent calculations. For ice VIII, the speedup ratio was even larger, close to 5:1.

\subsection{Solid hydrogen}

\begin{table*}[t]
\caption{\label{tab:hydrogen_geometry} Intramolecular H-H bond distance ($d_{\rm H-H}$) [\AA] and highest vibron frequency ($\nu_{\rm H-H}$) [cm$^{-1}$] in the $Cmca$-$4$ and $C2/c$-$24$ phases of solid hydrogen at 180~GPa, calculated with PBE, PBE0, PBE0[PBE], and the optimized hybrid functional with 48\% of exact exchange. The MA\%E of all optical $\Gamma$-point vibrational frequencies ($\nu$) (excluding the two imaginary modes for the $C2/c$-$24$ phase) is also given with respect to the scf PBE0/PBE0(48\%) hybrid calculation.
}
\setlength{\tabcolsep}{8pt}
\begin{tabular}{l|ccc|ccc}
\hline
\hline
    \textbf{Solid}  &  \multicolumn{3}{c|}{\textbf{$\boldsymbol{C2/c}$-$\boldsymbol{24}$}} & \multicolumn{3}{c}{\textbf{$\boldsymbol{Cmca}$-$\boldsymbol{4}$}} \\
\hline
    \textbf{[\AA]/[cm$^{-1}$]}  &  $d_{\rm H-H}$ & $\nu_{\rm H-H}$ & MA\%E ($\nu$) & $d_{\rm H-H}$ & $\nu_{\rm H-H}$ & MA\%E ($\nu$) \\
\hline
    \textbf{PBE}              &  0.7450 & 4316 & 3.52/6.28 & 0.7209 & 4683 & 6.42/10.54 \\
    \textbf{PBE0}             &  0.7266 & 4620 & 0.00      & 0.7098 & 4883 & 0.00       \\
    \textbf{PBE0[PBE]}        &  0.7267 & 4619 & 0.27      & 0.7095 & 4889 & 0.16       \\
    \textbf{PBE0(48\%)}       &  0.7138 & 4844 & 0.00      & 0.7013 & 5041 & 0.00       \\
    \textbf{PBE0(48\%)[PBE]}  &  0.7135 & 4852 & 0.73      & 0.7002 & 5063 & 0.43       \\
\hline
\end{tabular}
\end{table*}

In the range 100-300 GPa, hydrogen is known to form molecular solids \cite{gregoryanz2020EverythingYouAlways}. The various existing phases have, however, been difficult to characterize: experimentally due to the high pressures involved, and theoretically due to the crucial importance of electron-electron, electron-nuclei, and nuclei-nuclei interactions, that popular DFT functionals often fail to capture.  
For example, in order to obtain a reasonable description of the II-III phase transition with the PBE0 hybrid functional, the fraction of exact exchange needs to be increased significantly, from the standard 25\% to 48\% \cite{hellgren2022HighpressureIIIIIPhase, contant2024AssessingManybodyMethods}.

Here, we consider two candidate molecular phases of solid hydrogen, both predicted stable at 180 GPa \cite{pickard2007structure}. Following standard convention, they are denoted by their symmetry space group and number of atoms: $Cmca$-$4$ and $C2/c$-$24$. While $Cmca$-$4$ is a potential candidate for phase I or II, $C2/c$-$24$ reproduces, at least qualitatively, most experimental features of phase~III. In the case of $Cmca$-$4$, there are two different structures with the same symmetry. Below we consider its low-pressure variant, which is gapped.

The $Cmca$-$4$ and $C2/c$-$24$ structures are first fully relaxed at 180 GPa using the vdW-DF functional \cite{dion2004VanWaalsDensity,thonhauser2015SpinSignatureNonlocal}, which is found to be more accurate than PBE \cite{PhysRevB.89.184106,hellgren2022HighpressureIIIIIPhase}. Then, the atomic positions are relaxed using PBE, PBE0, and PBE0[PBE]. The optimized PBE0 hybrid functional with 48\% of exact exchange, denoted PBE0(48\%) hereafter, is also tested, both at self-consistency and starting from PBE orbitals (PBE0(48\%)\text{[}PBE\text{]}), in order to compare the accuracy of the nscf forces given an increased fraction of exact exchange $\a$. For $Cmca$-$4$, results converge within PBE0/PBE on a 10$\times$10$\times$10/17$\times$17$\times$17 $k$-point grid, and a kinetic energy cutoff of 80/120~Ry, respectively. 
For $C2/c$-$24$, we instead used 5$\times$5$\times$5/8$\times$8$\times$8 and 80/120~Ry for PBE0/PBE, respectively.

In Tab.~\ref{tab:hydrogen_geometry}, we show the H$_2$ intramolecular bond distance, $d_{\rm H-H}$, within the solids. The two H$_2$ molecules in $Cmca$-$4$ have the same bond distance due to symmetry, while in $C2/c$-$24$ there are three symmetry-inequivalent H$_2$ molecules. The molecule with the longest bond distance is shown in the Table. Due to the finite pressure, the H-H intramolecular bonds are all shorter in these two solids compared to the bond distance of an H$_{2}$ molecule at ambient conditions.

For $Cmca$-$4$, PBE0 predicts $d_{\rm H-H}=$ 0.7098~\AA, against 0.7095~\!\AA~for PBE0[PBE], a difference of only 0.0003~\!\AA. Using the larger fraction of exact exchange (48\%) increases this difference to 0.0011~\!\AA. This observation is expected, since a larger $\a$ makes the PBE starting point less suitable. It is, however, worth noting that this discrepancy remains a factor of almost 20 smaller than the difference observed with respect to PBE (0.7209~\AA).
Similar conclusions can be drawn for $C2/c$-$24$.

On top of the optimal geometry found for each method by the BFGS algorithm, we then computed the phonon frequencies at the $\Gamma$-point. Thanks to crystal symmetries, only 4 distorted geometries are needed in $Cmca$-$4$ to obtain the 12 vibrational modes. $C2/c$-$24$, instead, requires as many as 36 different distorted unit cells to obtain its 72 normal modes. 
Among the many phonons, those with a frequency higher than 4000~cm$^{-1}$ describe the vibration of the H$_{2}$ intramolecular bond, and are called vibrons. In Tab.~\ref{tab:hydrogen_geometry}, we list the most energetic of these frequencies. Due to the smaller H-H bond distances in the compressed solids, these frequencies are higher than the one of the isolated hydrogen molecule at zero pressure. If we compare the scf and nscf values, we see that with 25\% of exact exchange, the difference is only within 6~cm$^{-1}$. With 48\% of exact exchange, it reaches 22~cm$^{-1}$, or 0.4 \%.

The mean absolute percentage error (MA\%E) of all optical phonon modes at $\G$, calculated with respect to scf PBE0/PBE0(48\%), is also given in Tab~\ref{tab:hydrogen_geometry}.
For PBE0[PBE], this error is below 0.27\% on $C2/c$-$24$ and below 0.16\% on $Cmca$-$4$. For PBE0(48\%)[PBE], the error is below 0.73/0.43\% for $C2/c$-$24$/$Cmca$-$4$, respectively. On the other hand, for PBE, the differences are as large as 3.52/6.42\% with respect to PBE0 and 6.28/10.54\% to PBE0(48\%), respectively. To obtain these values for $C2/c$-$24$, we have had to omit the two negative (imaginary) frequencies, predicted by the hybrid functionals. These two phonon modes are predicted stable by PBE down to 120 GPa.
The finding of these imaginary modes using hybrid functionals agrees with Ref.~\cite{hellgren2022HighpressureIIIIIPhase}.
Despite the fact that these phonon instabilities are not identified by the PBE method, the use of PBE orbitals in nscf PBE0 calculations does not prevent their correct identification, although it has an impact on their calculated frequency (most negative mode is at $-$99~cm$^{-1}$/$-$62~cm$^{-1}$ in PBE0/PBE0[PBE], against $-$314~cm$^{-1}$/$-$299~cm$^{-1}$ in PBE0(48\%)/PBE0(48\%)[PBE]), which may slightly shift the predicted transition pressures.

In terms of computational cost, when ran on a 32-core local machine, the cost of a non-self-consistent calculation stays the same independently of the fraction of exact exchange used. Since $\a$ only acts as a multiplication factor in Eq.~(10), both nscf results can be obtained simultaneously, from the same calculation. In the case of self-consistent PBE0, the computational cost increases slightly with the fraction of exact exchange due to a harder convergence from the PBE starting point. For the smaller $Cmca$-$4$ phase, the speedup ratio observed when using nscf reaches 11:1 for PBE0 with the standard 25\% of exact exchange, and 15:1 with the exact exchange fraction set to 48\%. In the case of the larger $C2/c$-$24$ phase, as expected, a speedup is still observed but less pronounced: it approaches 2:1 for standard PBE0 and 3:1 for PBE0(48\%).

\subsection{SiO$_2$ and TiO$_2$}

\begin{table}[ht!]
\caption{\label{tab:SiO2_structure} Structural parameters and $\G$-point optical vibrational frequencies [cm$^{-1}$] of $\alpha$-SiO$_{2}$, calculated with PBE, PBE0, and PBE0[PBE]. For every method, the geometry is relaxed considering the experimental unit cell ($a\!=\!4.916$~\!\AA, \, $c\!=\!5.405$~\!\AA). The structural parameters are illustrated in Ref.~\cite{Levien1980}. The MA\%E ($\nu$) of optical frequencies with respect to scf PBE0 is also given. The phonon frequencies affected by the LO-TO splitting (not considered in the calculations) are underlined. Experimental results are from Ref.~\cite{Levien1980}.
}
\centering
\setlength{\tabcolsep}{3pt}
\begin{tabular}{lccccc}
\hline
                    &  \textbf{PBE} & \textbf{PBE0} & \textbf{PBE0[PBE]} & \textbf{Expt.} \\
\hline
    \textbf{x(Si)}  &  0.4685 & 0.4727 & 0.4729 & 0.4697 \\
    \textbf{x(O)}   &  0.4120 & 0.4147 & 0.4148 & 0.4135 \\
    \textbf{y(O)}   &  0.2694 & 0.2623 & 0.2619 & 0.2669 \\
    \textbf{z(O)}   &  0.1172 & 0.1235 & 0.1238 & 0.1191 \\
    \textbf{$\boldsymbol{d}_{\text{Si-O}}$ \,[\AA]}  &  1.612 & 1.599 & 1.599 & 1.605 \\
                                        &  1.617 & 1.603 & 1.602 & 1.614 \\
    \textbf{$\boldsymbol{\Theta}_{\text{Si-O-Si}}$ \,[$\boldsymbol{^{\circ}}$]}  &  142.7 & 145.3 & 145.4 & 143.7 \\
    \textbf{$\boldsymbol{\Theta}_{\text{O-Si-O}}$ \,[$\boldsymbol{^{\circ}}$]}   &  108.7 & 108.9 & 108.9 & 108.8 \\
                                                                                 &  108.7 & 108.9 & 108.9 & 109.0 \\
                                                                                 &  109.1 & 109.2 & 109.2 & 109.2 \\
                                                                                 &  110.8 & 110.4 & 110.4 & 110.5 \\
\\
    \textbf{E$_{\text{u}}$(TO1)}  &            \,\,140  &            \,\,\,\,67  &            \,\,\,\,76  & \,\,133 \\
    \textbf{E$_{\text{u}}$(LO1)}  & \,\,\underline{140} & \,\,\,\,\underline{67} & \,\,\,\,\underline{76} & \,\,133 \\
    \textbf{A$_{\text{1}}$(1)}    &            \,\,211  &               \,\,167  &               \,\,172  & \,\,219 \\
    \textbf{E$_{\text{u}}$(TO2)}  &            \,\,263  &               \,\,242  &               \,\,243  & \,\,269 \\
    \textbf{E$_{\text{u}}$(LO2)}  & \,\,\underline{263} &    \,\,\underline{242} &    \,\,\underline{243} & \,\,269 \\
    \textbf{A$_{\text{1}}$(2)}    &            \,\,351  &               \,\,354  &               \,\,355  & \,\,358 \\
    \textbf{A$_{\text{2}}$(TO1)}  &            \,\,352  &               \,\,363  &               \,\,366  & \,\,361 \\
    \textbf{E$_{\text{u}}$(TO3)}  &            \,\,386  &               \,\,388  &               \,\,390  & \,\,394 \\
    \textbf{E$_{\text{u}}$(LO3)}  & \,\,\underline{386} &    \,\,\underline{388} &    \,\,\underline{390} & \,\,402 \\
    \textbf{E$_{\text{u}}$(TO4)}  &            \,\,442  &               \,\,446  &               \,\,448  & \,\,453 \\
    \textbf{E$_{\text{u}}$(LO4)}  & \,\,\underline{442} &    \,\,\underline{446} &    \,\,\underline{448} & \,\,512 \\
    \textbf{A$_{\text{1}}$(3)}    &            \,\,452  &               \,\,453  &               \,\,454  & \,\,469 \\
    \textbf{A$_{\text{2}}$(TO2)}  &            \,\,484  &               \,\,489  &               \,\,491  & \,\,499 \\
    \textbf{E$_{\text{u}}$(TO5)}  &            \,\,677  &               \,\,696  &               \,\,698  & \,\,698 \\
    \textbf{E$_{\text{u}}$(LO5)}  & \,\,\underline{677} &    \,\,\underline{696} &    \,\,\underline{698} & \,\,701 \\
    \textbf{A$_{\text{2}}$(TO3)}  &            \,\,756  &               \,\,782  &               \,\,784  & \,\,778 \\
    \textbf{E$_{\text{u}}$(TO6)}  &            \,\,776  &               \,\,797  &               \,\,798  & \,\,799 \\
    \textbf{E$_{\text{u}}$(LO6)}  & \,\,\underline{776} &    \,\,\underline{797} &    \,\,\underline{798} & \,\,812 \\
    \textbf{E$_{\text{u}}$(TO7)}  &               1042  &                  1085  &                  1088  &    1066 \\
    \textbf{E$_{\text{u}}$(LO7)}  &    \underline{1042} &       \underline{1085} &       \underline{1088} &    1227 \\
    \textbf{A$_{\text{2}}$(TO4)}  &               1050  &                  1092  &                  1095  &    1072 \\
    \textbf{A$_{\text{1}}$(4)}    &               1061  &                  1101  &                  1105  &    1082 \\
    \textbf{E$_{\text{u}}$(TO8)}  &               1133  &                  1181  &                  1185  &    1158 \\
    \textbf{E$_{\text{u}}$(LO8)}  &    \underline{1133} &       \underline{1181} &       \underline{1185} &    1155 \\
\\[-10pt]
    \textbf{MA\%E ($\nu$)}        &              12.80  &                  0.00  &                  1.55  &     /   \\
\hline
\end{tabular}
\end{table}

\begin{table}[t]
\caption{\label{tab:tio2_structure} Lattice parameters and $\G$-point optical vibrational frequencies [cm$^{-1}$] of the rutile phase of TiO$_{2}$, calculated with PBE, PBE0, and PBE0[PBE]. The MA\%E ($\nu$) of optical phonon frequencies is also given with respect to scf PBE0 (excluding the A$_{\text{2u}}$ mode). The phonon frequencies affected by the LO-TO splitting (not considered in the calculations) are underlined.
}
\centering
\setlength{\tabcolsep}{6pt}
\begin{tabular}{lccc}
\hline
                        &  \textbf{PBE} & \textbf{PBE0} & \textbf{PBE0[PBE]} \\
\hline
    \textbf{$\boldsymbol{a}$ \,[\AA]}  &  4.642  & 4.581  & 4.582  \\
    \textbf{$\boldsymbol{c}$ \,[\AA]}  &  2.968  & 2.948  & 2.944  \\
    \textbf{$\boldsymbol{u}$}          &  0.3050 & 0.3052 & 0.3054 \\
\\
    \textbf{B$_{\text{1u}}$(1)}   &             \,\,\,28  &            \,\,\,77  &             \,\,\,49  \\
    \textbf{A$_{\text{2u}}$}      & \!\!\underline{$-$89} & \,\,\,\underline{83} & \!\!\underline{$-$37} \\
    \textbf{E$_{\text{u}}$(TO1)}  &             \,\,\,15  &                 100  &             \,\,\,62  \\
    \textbf{E$_{\text{u}}$(LO1)}  &    \,\,\,\underline{15} &      \underline{100} &  \,\,\,\underline{62} \\
    \textbf{B$_{\text{1g}}$}      &                  166  &                 173  &                  173  \\
    \textbf{E$_{\text{u}}$(TO2)}  &                  361  &                 384  &                  383  \\
    \textbf{E$_{\text{u}}$(LO2)}  &       \underline{361} &      \underline{384} &       \underline{383} \\
    \textbf{B$_{\text{1u}}$(2)}   &                  353  &                 391  &                  387  \\
    \textbf{A$_{\text{2g}}$}      &                  425  &                 454  &                  454  \\
    \textbf{E$_{\text{g}}$(1)}    &                  426  &                 459  &                  459  \\
    \textbf{E$_{\text{g}}$(2)}    &                  426  &                 459  &                  459  \\
    \textbf{E$_{\text{u}}$(TO3)}  &                  474  &                 512  &                  512  \\
    \textbf{E$_{\text{u}}$(LO3)}  &       \underline{474} &      \underline{512} &       \underline{512} \\
    \textbf{A$_{\text{1g}}$}      &                  577  &                 614  &                  614  \\
    \textbf{B$_{\text{2g}}$}      &                  778  &                 836  &                  839  \\
\\[-10pt]
    \textbf{MA\%E ($\nu$)}     &                30.48  &                0.00  &                 8.18  \\
    \textbf{MA\%E ($\nu>100$)}    &                 6.78  &                0.00  &                 0.23  \\
\hline
\end{tabular}
\vspace{1pt}
\end{table}

In this Subsection, we consider the $\alpha$-quartz phase of SiO$_{2}$ ($P3_{2}21$, space group no.~\!154) and the rutile phase of TiO$_{2}$ ($P4_{2}/mnm$, space group no.~\!136). The structure of TiO$_{2}$ is determined by the lattice parameters $a$ and $c$ of the tetragonal crystal system, along with a single internal coordinate $u$ that fixes the geometry of the oxygen atoms in the primitive cell \cite{howard1991StructuralThermalParameters}. The structure of SiO$_{2}$ is determined by the lattice parameters $a$ and $c$ of the trigonal crystal system, and its geometry is determined by a set of parameters, illustrated in Ref.~\cite{Levien1980}.
Similarly to previous systems, we compare the geometry and $\G$-point phonon frequencies of PBE, PBE0, and PBE0[PBE].

To ensure converged results for the structure and geometry optimization, a kinetic energy cutoff of 100/120~Ry and a 3$\times$3$\times$3/5$\times$5$\times$5 uniform $k$-point grid was used for SiO$_2$/TiO$_2$, respectively. Because SiO$_{2}$ already has many internal geometry degrees of freedom, we considered this material at fixed lattice parameters, set to the experimental values ($a\!=\!4.916$~\!\AA, \, $c\!=\!5.405$~\!\AA)~\cite{Levien1980}. However, because TiO$_{2}$ only has a single internal degree of freedom to define every atomic position within its unit cell, we decided to fully relax, for each method, both its geometry and lattice parameters, in order to compare how different the optimal structures look like. Results can be found in Tab.~\ref{tab:SiO2_structure} for SiO$_{2}$ and Tab.~\ref{tab:tio2_structure} for TiO$_{2}$.

Because both materials are polar, the LO-TO splitting should also be considered \cite{shojaee2009FirstprinciplesElasticThermal}. However, since we are mainly interested in comparing the performance of PBE0[PBE] with respect to scf PBE0, this shift has not been calculated. The phonon frequencies affected have been underlined in the two Tables. For TiO$_{2}$, we found that the convergence of the PBE vibrational frequencies is slow with respect to both $k$-points and cutoff. As such, for PBE, we used in Tab.~\ref{tab:tio2_structure} an 8$\times$8$\times$8 uniform $k$-point grid and a 200~Ry plane-wave cutoff.

In Ref.~\cite{contant2024OptimizedEffectivePotential}, SiO$_{2}$ was used as a test system to assess the accuracy of forces obtained with the PBE0 method as implemented within the OEP framework. We can now extend this analysis by evaluating the performance of using an "unoptimized" KS starting point as in PBE0[PBE]. Looking at the results in Tab.~\ref{tab:SiO2_structure} for SiO$_{2}$, we find a geometry with PBE0[PBE] that is in very good agreement with PBE0. The maximum difference observed across every bond angle is within 0.1$^{\circ}$. The difference in Si-O bond distances is also minimal, within 0.001~\!\AA. On these optimized geometries, we then computed the phonon modes at the $\Gamma$-point. The different optical frequencies are listed in Tab.~\ref{tab:SiO2_structure}.

Compared to OEP-PBE0 (see Ref.~\cite{contant2024OptimizedEffectivePotential}), we notice a larger difference between the self-consistent and non-self-consistent calculations at the PBE0 level. These differences appear to mostly affect the three lowest phonon frequencies and are within 10~cm$^{-1}$. For the other phonon modes, the differences are smaller, e.g., within 2~cm$^{-1}$ for the A$_{2}$(TO2) mode around 490~cm$^{-1}$, and within 4~cm$^{-1}$ for the A$_{1}$(4) mode located at almost twice that frequency. The MA\%E of all frequencies is 1.55\%. However, excluding the three least energetic phonon modes reduces the MA\%E to 0.37\%. Thus, overall, the level of description provided by PBE0[PBE] is very good, not only in predicting the geometry but also for performing the vibrational analysis.

For TiO$_2$, as mentioned before, the structure was fully relaxed with PBE, PBE0, and PBE0[PBE]. Comparing PBE and PBE0 (see Tab.~\ref{tab:tio2_structure}), the largest difference is seen in the lattice parameter $a$. PBE0 compresses the structure, reducing the error with respect to experiment ($a\!=\!4.587$~\!\AA, $c\!=\!2.954$~\!\AA ~\cite{burdett1987StructuralelectronicRelationshipsInorganic}) from 1.20\% to 0.13\%. PBE0[PBE] behaves similarly (0.11\% error with respect to experiment). On the other hand, for the lattice parameter $c$, the error with respect to experiment is only reduced slightly by the hybrid functionals, from 0.47\% in PBE to 0.20\% in PBE0 and 0.34\% in PBE0[PBE].

\begin{table*}[t]
\caption{\label{tab:csibn} Structural parameters and $\G$-point ZO and LO/TO vibrational frequencies [cm$\boldsymbol{^{-1}}$] of monolayer C, Si, and BN, calculated with PBE, PBE0, and PBE0[PBE]. The structural parameters of silicene are illustrated in Ref.~\cite{ZHAO201624}.
}
\setlength{\tabcolsep}{5.5pt}
\begin{tabular}{l|ccc|cccccc|ccc}
\hline
\hline
                        &  \multicolumn{3}{c|}{\textbf{Graphene}} & \multicolumn{6}{c|}{\textbf{Silicene}} & \multicolumn{3}{c}{\textbf{2D-BN}} \\
\hline
    \textbf{[\AA]/[$\boldsymbol{^{\circ}}$]/[cm$\boldsymbol{^{-1}}$]}  & $a$ & ZO & LO/TO & $a$ & $d_{\rm Si-Si}$ & $\theta$ & $h$ & ZO & LO/TO & $a$ & ZO & LO/TO \\
    \hline
    \textbf{PBE}        &  2.4630 & 878 & 1556 &  3.8716 & 2.2834 & 115.94 & 0.4666 & 191 & 548 & 2.5073 & 803 & 1341 \\
    \textbf{PBE0}       &  2.4474 & 917 & 1621 &  3.8540 & 2.2604 & 116.97 & 0.3978 & 165 & 581 & 2.4914 & 840 & 1395 \\
    \textbf{PBE0[PBE]}  &  2.4466 & 918 & 1625 &  3.8508 & 2.2598 & 116.87 & 0.4046 & 168 & 583 & 2.4908 & 841 & 1397 \\
    \hline
\end{tabular}
\end{table*}

There are, in total, 15 optical modes in TiO$_2$. The symmetry of each normal mode has been identified using Ref.~\cite{montanari2002LatticeDynamicsTiO2}. The frequency associated with the A$_{\text{2u}}$ mode is found negative (or imaginary) in PBE, which has already been established in previous works \cite{montanari2002LatticeDynamicsTiO2,shojaee2009FirstprinciplesElasticThermal}. The PBE method thus tends to favor the formation of a ferroelectric material, which is an incorrect description for the rutile phase of TiO$_2$ \cite{montanari2002LatticeDynamicsTiO2}. The frequencies predicted by PBE0 are, as expected, larger than with PBE. No imaginary mode is observed. If we compare with PBE0[PBE], we see that, for most phonon modes, the agreement holds well, with differences not exceeding 4~cm$^{-1}$. Nevertheless, disagreements start to become visible when considering the four lowest frequencies. PBE0[PBE] inherits some of the deficiencies of PBE. Similarly to PBE, the A$_{\text{2u}}$ mode remains unstable, unlike scf PBE0. This corresponds to a case where PBE orbitals are not sufficient for PBE0. The difference for the other three optical modes with lowest frequency (B$_{\text{1u}}$ and the two E$_{\text{u}}(1)$ modes) are also quite large, about 30-40~cm$^{-1}$. Nevertheless, PBE0[PBE] is still able to improve upon PBE, shifting the results by 20-50~cm$^{-1}$.

In terms of computational cost, when ran on a local 32-core machine, the non-self-consistent calculations performed on TiO$_{2}$ offered a speedup ratio of 4:1 over scf PBE0. The nscf approach is, therefore, still relevant if the interest lies in finding the optimal geometry and evaluating the vibrational modes above 100~cm$^{-1}$, for which the accuracy matches the fully self-consistent results. The nscf calculation for SiO$_{2}$ ends up being 2 times faster than scf PBE0.

\begin{table}[t]
\centering
\caption{Structural parameters and optical vibrational frequencies [cm$^{-1}$] at the $\Gamma$-point of phosphorene, calculated with PBE, PBE0, and PBE0[PBE]. The structural parameters are illustrated in Ref.~\cite{doi:10.1143/JPSJ.50.3362}. The MA\%E ($\nu$) of optical phonon frequencies is also given with respect to scf PBE0.
}
\label{tab:P4_structure}
\setlength{\tabcolsep}{8pt}
\begin{tabular}{lccc}
\hline
                              & \textbf{PBE} & \textbf{PBE0} & \textbf{PBE0[PBE]} \\
\hline
    \textbf{$\boldsymbol{a}$ \,[\AA]}        &     4.624    &      4.582    &           4.579    \\
    \textbf{$\boldsymbol{b}$ \,[\AA]}        &     3.305    &      3.284    &           3.282    \\
    \textbf{$\boldsymbol{u}$}                &     0.08871  &      0.08969  &           0.08958  \\
    \textbf{$\boldsymbol{v}$}                &     0.08664  &      0.08535  &           0.08536  \\
    \textbf{$\boldsymbol{d_{1}}$ \,[\AA]}  &     2.226    &      2.203    &           2.203    \\
    \textbf{$\boldsymbol{d_{2}}$ \,[\AA]}  &     2.263    &      2.234    &           2.234    \\
    \textbf{$\boldsymbol{\theta_{1}}$ \,[$\boldsymbol{^{\circ}}$]}  &   95.85 &  96.37 &  96.33 \\
    \textbf{$\boldsymbol{\theta_{2}}$ \,[$\boldsymbol{^{\circ}}$]}  &  104.06 & 104.20 & 104.18 \\
\\
    \textbf{B$_{\text{3u}}$}  &  138 &  153 &  154 \\
    \textbf{B$_{\text{3g}}$}  &  187 &  202 &  204 \\
    \textbf{B$_{\text{2g}}$}  &  222 &  240 &  241 \\
    \textbf{A$_{\text{g}}$}   &  344 &  369 &  369 \\
    \textbf{A$_{\text{u}}$}   &  414 &  453 &  454 \\
    \textbf{B$_{\text{1g}}$}  &  421 &  462 &  463 \\
    \textbf{B$_{\text{2g}}$}  &  426 &  463 &  464 \\
    \textbf{A$_{\text{g}}$}   &  451 &  487 &  488 \\
    \textbf{B$_{\text{1u}}$}  &  458 &  494 &  494 \\
\\[-10pt]
    \textbf{MA\%E ($\nu$)}    & 7.97 & 0.00 & 0.34 \\
\hline
\end{tabular}
\end{table}

\subsection{Monolayer C, Si, BN, and P}

Finally, we consider a set of monolayer materials: graphene, silicene, boron nitride, and phosphorene. For each system, the geometry is optimized with PBE, PBE0, and PBE0[PBE]. While graphene and monolayer BN (2D-BN) relax in perfectly flat 2D hexagonal layers, silicene admits a buckling. As such, two additional structural parameters are introduced: the deviation from the flat plane, "$h$", and the angle, $\theta$, of the distorted hexagon \cite{ZHAO201624}, now smaller than 120$^{\circ}$. Phosphorene corresponds to a single layer of black phosphorus, which crystallizes in the orthorhombic system ($Cmce$, space group no.~\!64). The set of parameters determining the geometry of the P layer is illustrated in Refs.~\cite{doi:10.1143/JPSJ.50.3362,carvalho2016PhosphoreneTheoryApplications}. 

Graphene and silicene are semi-metals, with a Dirac cone at the high-symmetry point $K$. A dense $k$-point sampling is, therefore, necessary. Indeed, considering a Fermi-Dirac smearing with a spreading of 0.0150~Ry for graphene and 0.0075~Ry for silicene, results converge with a 24$\times$24$\times$1/18$\times$18$\times$1 $k$-point grid for monolayer C and Si, respectively. A plane-wave cutoff of 120 Ry for graphene and 80 Ry for silicene is used.
Monolayer BN and P are gapped and converge with a 120/80~Ry plane-wave cutoff and a 9$\times$9$\times$1/8$\times$8$\times$1 $k$-point grid, respectively. A vacuum layer of 12~\AA~is used for every system. The optimized structural parameters and the $\Gamma$-point frequencies are listed in Tab.~\ref{tab:csibn} (graphene, silicene, and 2D-BN) and Tab.~\ref{tab:P4_structure}~(phosphorene).

The structure of graphene and 2D-BN are determined solely by their lattice parameter. As compared to PBE, PBE0 compresses the structures by 0.6\% and increases the ZO and LO/TO vibrational frequencies by about 4\%. The difference between PBE0[PBE] and PBE0 is instead as small as 0.03\% for the lattice constants, and 0.2\% for the optical phonon frequencies. In general, phonon frequencies harden going from PBE to PBE0. An exception is the ZO mode of silicene, which softens by 13\% with PBE0. This is most likely due to the reduced buckling, described by the $h$-parameter. Despite the slightly more complex structure, the differences between PBE0 and PBE0[PBE] are very small also in silicene, comparable to the differences found in both graphene and 2D-BN.

\begin{figure}[ht!]
\centering
\includegraphics[scale=0.56]{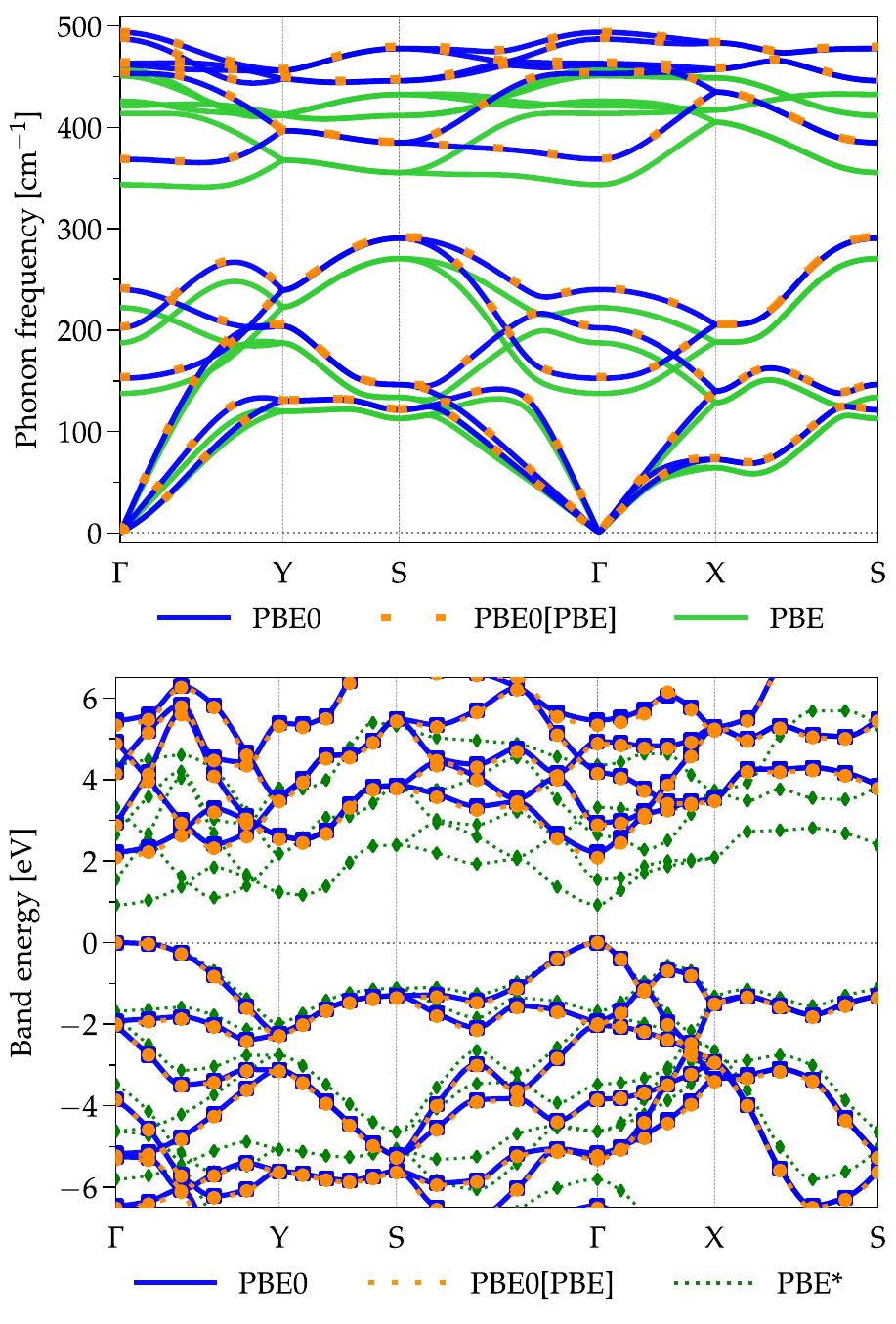}
\caption{Phonon dispersion (top) and electronic band structure (bottom) of phosphorene, calculated with PBE0 (blue), PBE0[PBE] (orange dots), and PBE (green), on top of their respective optimal geometry (see Tab.~\ref{tab:P4_structure}). For PBE* (green dashes), the PBE0[PBE] optimal geometry was used.
}\label{fig:phosphorene}
\end{figure}

Similar conclusions can be drawn for phosphorene, as shown in Tab.~\ref{tab:P4_structure}. The optimal PBE0[PBE] geometry is almost identical to the PBE0 geometry. The phonon frequencies at the $\G$-point are also nearly indistinguishable. Among the 9 optical modes, the maximal frequency difference observed does not exceed 2~cm$^{-1}$. This good agreement also extends to the phonon dispersion, obtained on a [2,2,1] supercell (16 atoms) and shown in the top panel of Fig.~\ref{fig:phosphorene}.

To determine whether PBE0 and PBE0[PBE] also agree on electronic spectral properties, we proceeded to calculate the electronic band structure of phosphorene. The eigenvalues are calculated on a 10$\times$10$\times$1 $k$-point grid. These values are represented as data points in the bottom panel of Fig.~\ref{fig:phosphorene}. To recover a smooth band description along the selected $k$-point path, these data points have been spline-interpolated. The uncorrected set of eigenvalues (shown in dark green) corresponds to PBE but calculated on the PBE0[PBE] geometry (marked as PBE* to avoid any confusion). The role of self-consistency in PBE0 is minimal for the calculation of the band structure, in accordance with the findings of Ref.~\cite{skelton2020}. The PBE0 and PBE0[PBE] results are in very good agreement, especially for the valence bands.

\begin{figure}[ht!]
\centering
\includegraphics[scale=0.36]{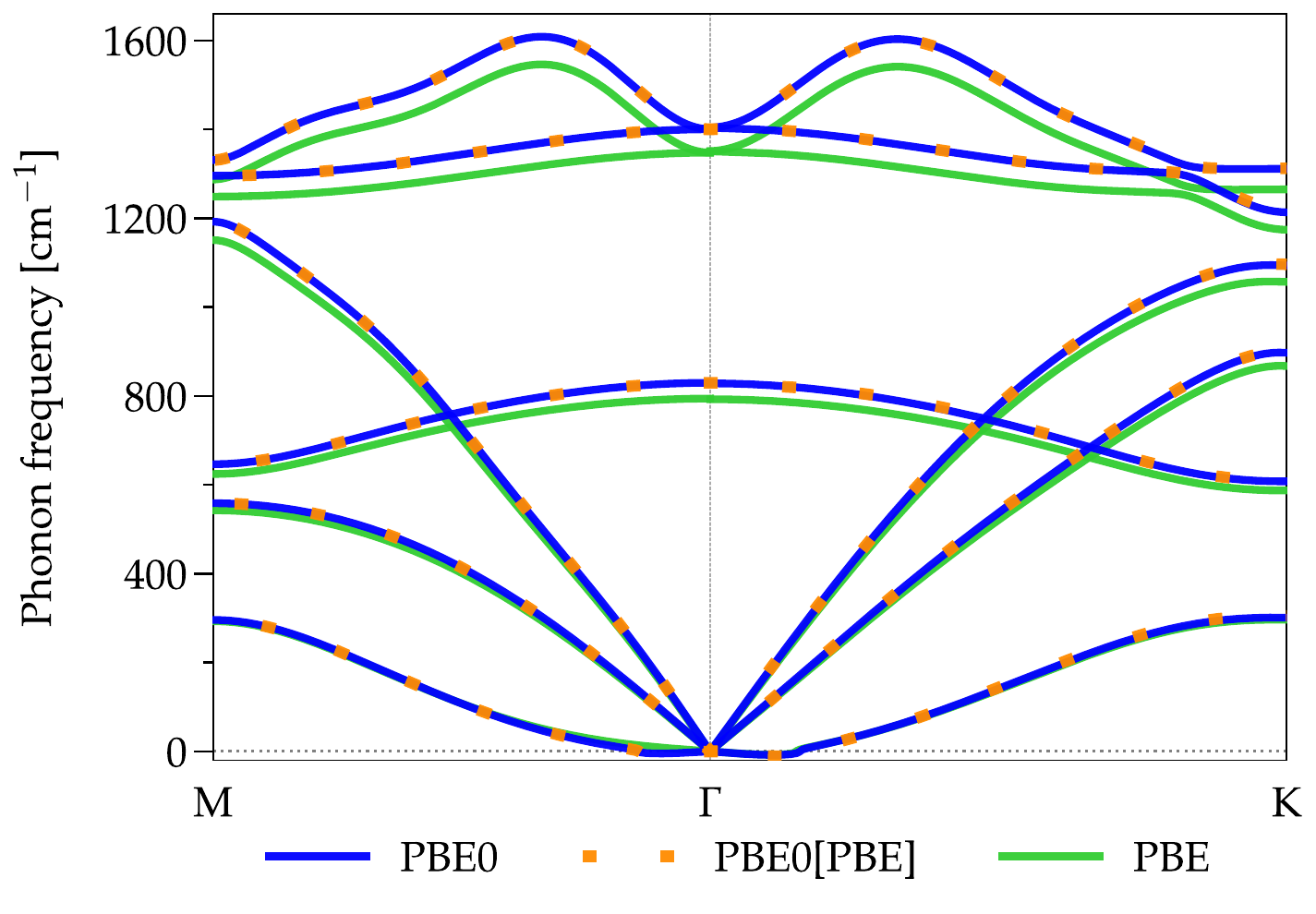}
\caption{Phonon dispersion of 2D-BN, calculated with PBE0 (blue), PBE0[PBE] (orange dots), and PBE (green), on top of their respective optimal geometry (see Tab.~\ref{tab:csibn}).}
\label{fig:2dhBN_dispersion}
\end{figure}

In addition to phosphorene, we also evaluated the phonon dispersion of 2D-BN. It is shown in Fig.~\ref{fig:2dhBN_dispersion}. The $\Gamma-$M and $\Gamma-$K paths have been obtained by considering two different supercells. For the first path, a [8,1,1] supercell (16 atoms) has been used in order to capture four intermediate points accurately, in addition to $\Gamma$ and M. To obtain the same number of accurate points along $\Gamma-$K, we used a [12,1,1] supercell (24 atoms) built upon a reoriented primitive unit cell that aligns point K along a single lattice vector. For both paths, we considered a 200~Ry kinetic energy cutoff, as we found the ZA branch to converge slowly with respect to the cutoff energy, for all three functionals considered. For the $k$-point sampling of the Brillouin zone, we used 2$\times$12$\times$1 and 1$\times$12$\times$1 for $\Gamma-$M and $\Gamma-$K, respectively.
These two phonon paths have been merged at the $\Gamma$-point (see Fig.~\ref{fig:2dhBN_dispersion}). An excellent agreement can once again be seen between PBE0 and PBE0[PBE].

Regarding the computational efficiency of the non-self-consistent force calculations, we are again able to notice a significant speedup for the $\Gamma$-point calculations. Since the unit cells are small and the number of $k$-points large, the DFPT step is very quick in comparison to the evaluation of the Fock operator. On a local machine with 32 cores, the speedup ratio observed with PBE0[PBE] exceeds 9:1 in the case of graphene. For silicene, we find 10:1. For 2D-BN and phosphorene, the speedup is also important, of the order of 4:1.

In the case of the supercells considered for phosphorene and 2D-BN, the speedup advantage of nscf becomes less important due to the increase in system size and the reduction in the number of sampled $k$-points. For phosphorene, the nscf calculations done on the [2,2,1] distorted supercells are found to be 2 times faster than the fully self-consistent PBE0 calculations, ran on a supercomputer with 32 cores. For 2D-BN, the nscf and scf approaches end up being almost equivalent.

\section{Conclusions}

In this work, we have investigated the effects of self-consistency when evaluating atomic forces with hybrid functionals. Non-self-consistent forces necessitate the evaluation of extra force terms that involve the calculation of self-consistent orbital responses with respect to atomic displacements. Such responses have, here, been evaluated using the DFPT framework, as implemented within the \verb|Quantum ESPRESSO| distribution.

Tests performed on a number of systems show that geometries and vibrational frequencies across the Brillouin zone are all well-reproduced by non-self-consistent calculations. The structure and geometry are essentially identical to those obtained in a fully self-consistent calculation, and phonon frequencies are typically agreeing within 0.5\%, with only a few exceptions seen for low-energy modes and where the semi-local starting point has structural instabilities, absent from the hybrid functional.

Throughout this work, we have considered the PBE0 hybrid functional with the PBE starting point. This nscf approach can, naturally, be generalized to any fraction of exact exchange, $\a$, as well as to any type of hybrid functional. Looking at hydrogen solids, we found that increasing $\a$ beyond the standard 25\%, to 48\%, only marginally affected the nscf predictions. It is also interesting to note that since $\a$ only acts as a multiplication factor, the results from any fraction can be obtained in a single calculation.

The speedup observed with nscf calculations can be significant, up to a factor of 10 for small unit cells. Given the accuracy achieved for the phonon dispersion, it is, in the future, motivated to determine the vibrational frequencies through analytic second-order derivatives at the non-self-consistent hybrid level, which would bypass the use of large supercells.
\acknowledgements
The work was performed using high-performance computing resources from GENCI-TGCC/CINES/IDRIS (Grant No. A0150914650).\\

\providecommand{\noopsort}[1]{}\providecommand{\singleletter}[1]{#1}

\end{document}